**Ultra-small (r<2 nm), stable (>1 year), mixed valence copper oxide quantum dots with anomalous band gap**

**Bhusankar Talluri** [1,2], **Edamana Prasad**[2] **and Tiju Thomas** *,[1]

[1] Department of Metallurgical and Material Engineering, Indian Institute of Technology Madras, Chennai-600036, India

[2] Department of chemistry, Indian Institute of Technology Madras, Chennai-600036, India

Correspondence:

*,[1] E-mail: tijuthomas@iitm.ac.in, tt332@cornell.edu; Fax: +91-44- 2257-4752 ; Tel: +91- 44- 2257-5781 (Lab)

**ABSTRACT**

Ultra-small (r<2 nm) semiconductor quantum dots (QDs) have attracted attention for applications ranging from dye sensitized solar cells to sensing due to its tunable electronic structure and band gap, and large specific surface area. However obtaining monodisperse QDs and stabilization in this size regime remains a challenge. A recent report on digestive ripening of an oxide system showed substantial promise in addressing these requirements of QDs. In this work, we report a green solution, soft chemical (*chimie douce*) approach for synthesis of quasi-spherical, ultra-small, stable, and monodispersed copper oxide QDs (r<2 nm) based on digestive ripening (DR). It may be noted this is only the second transition metal oxide system in which DR is reported so far. DR involves the refluxing of polydispersed colloidal nanoparticles in the presence of surface active agents (e.g. triethanolamine (TEA)) that leads to fairly monodispersed nanoparticles. It has been noticed that capping with TEA results in reduction in the average particle diameter from 6 ± 4 nm to 2.4 ± 0.5 nm and an increase of zeta potential (ξ) from +12±2 mV to +31±2 mV. These copper oxide QDs are monodispersed (size ~ 2.4±0.5 nm), and stable (>1 year). In addition, these quantum dots show an anomalous increase in band gap (5.3 eV), inexplicable using Brus' equation. XPS indicates that Cu exists in mixed valence state in this material. Based on our observations, we suggest that off-stoichiometry in copper oxide, which is seemingly substantial at these length scales, is responsible for the observed anomaly in band gap.

1. Introduction

Semiconductor nanocrystals/quantum dots (QDs) have attracted considerable attention due to unique properties seen in them owing to quantum confinement, and high surface-to-volume ratios in these systems. However obtaining monodisperse QDs through chemical approaches remains a challenge [1]. QDs have already found applications in various areas of technology including photovoltaics [2–4], optoelectronics [5], catalysis [6], light-emitting diodes [7,8], biological fluorescent labels [9–11], solar cells [12], light harvesting [13], lasers [14], sensors [15], and controlled drug delivery [16]. Synthesis of QDs in the ultra-small size region (r<2) is a difficult task due to issues associated with controlling particle size and stabilization, both of which are necessary for practical applications. However recently ZnO ultra small QDs have been reported by Niya et al [17]. This was the first report on digestive ripening in an oxide ceramic system. Considering the relevance of oxide semiconductors, a reasonable question to ask would concern generalizing this result to other oxides.

Among various types of semiconductor nanocrystals, copper oxide semiconductors are attractive for both fundamental investigations and practical applications. In fact there have been reports wherein copper oxides has been explored for light harvesting [18]. Oxides of copper have rich surface chemistry and fairly specific chemical stability zones [19]. This combination makes chemical questions concerning copper oxides, in the ultra-small regime very interesting. On the applications front, so far copper oxide nanoparticles (sizes: 5-50 nm) have been used for applications such as solarcells [20], gas sensors [21], bio-sensors [22], nano fluids [23], photodetector [24], and photocatalysis [25,26]. In the ultra-small size form (r<2), copper oxide is expected to find use in photofunctional and optoelectronic applications (e.g. photo voltaic) [27,28]. One cannot discount applications in catalysis as well, since surface of certain oxides of copper can be metallic [29,30]. Prior to this work, copper oxide QDs of radius ~4-5 nm have been reported by Yee-Fun Lim et al.'s work based on solution approach [28]. In this report however obtaining monodispersity is an issue. In this work, following recent reports by Niya Mary et al.'s work [17], DR is explored as a possible route to obtain monodisperse copper oxide QDs.

Our studies have resulted in a green solution, rapid (~5 mins), approach for syntheses of ultra-small, stable (>1 yr), monodisperse (indicating DR), mixed valence copper oxide QDs having radius of less than 2 nm. The copper oxide QDs obtained thus are solution process able, stable (life time >1 year), inexpensive, and monodisperse. Furthermore, the anomalously large band gap these particles show (~5.3 eV) point to changes in the stoichiometry and electronic structure in these particles; the physics involved here merits further investigation.

## 2. Experimental section

### 2.1. Materials and method

For the synthesis reported here, the following chemicals are used without purification: copper acetate monohydrate $Cu(OAc)_2 \cdot H_2O$ (98%), ethanol (99.5%) and triethanolamine (TEA) (99.0%), sodium hydroxide (NaOH). $Cu(OAc)_2 \cdot H_2O$ and NaOH are the starting precursors for QD synthesis.

The reaction is carried out in ethanol as the medium. Ethanolic solution with concentration 3 mM of $Cu(OAc)_2 \cdot H_2O$ is prepared. Copper oxide synthesis is aided by the addition of NaOH (solid) to the reaction mixture at room temperature. A constant molar ratio of $Cu(OAc)_2 \cdot H_2O$ to NaOH is maintained (1:6); this ensures the formation of quasi-spherical particles and eliminates the possibility of habit development in QDs [31]. The mixture is stirred using a magnetic stirrer at 1100–1200 rpm. Once complete dissolution of NaOH, TEA is added to the ethanolic mixture; this must be done in a drop wise manner. TEA:$Cu(OAc)_2 \cdot 2H_2O$ molar ratio is varied systematically (1:1, 3:1, 5:1, 10:1 and 15:1), so as to study the effect of capping agent on copper oxide QDs. We observe that the most stable and smallest particles are formed when ratio is 10:1

*2.2. Characterization*

The crystal structure of as-synthesized copper oxide are recorded using X-ray diffraction (XRD) (PAN analytical X'pert PRO) using Cu K$_\alpha$ (wavelength 1.54 Å) source. In order to get sufficient sample for performing XRD, higher concentration of precursors (30 mM Cu(OAc)$_2$.H$_2$O and NaOH; and a molar ratio Cu(OAc)$_2$.H$_2$O:NaOH=1:6) is chosen. The as-obtained copper oxide powder is annealed at 300°C for 3 hours. Ultraviolet and visible absorption (UV– vis) spectrum is recorded with a Thermo evolution 201 spectrophotometer. Transmission Electron Micrographs (TEM images) are obtained from a Philips FEI TECNAI G2S TEM operating at 200 kV. X-ray photoelectron spectroscopy (XPS) measurements are carried out in a Omicron ESCA Probe spectrometer with polychromatic Mg Kα X-rays (hν = 1253.6 eV). The zetapotential (ξ) of QDs are measured using Malvern Zetasizer Nano ZS-90 instrument. FT-IR spectra are recorded using a Jasco FTIR-4100 spectrometer. Photoluminescence spectra are obtained by using Jobin Yvon Flurolog-3-11 spectrofluorimeter.

**3. Results and discussion**

The formation of CuO in bulk form from copper acetate as starting precursor is confirmed by XRD as shown in figure 1. Here the surface active agent has not been used. XRD clearly shows the presence of CuO phase with monoclinic crystal structure; this is in agreement with the data of ICSD No. 98-001-7494. We have observed that whenever surface active agent is used ultra-small copper oxide QDs formed. XPS data suggest that Cu exists in mixed valence (+1 and +2) states; since the QDs are sufficiently small, it is reasonable to believe that this is the case through the material. FTIR and PL suggest that presence mixed phases in these ultra-small copper oxide particles. However despite of presence mixed valencies, it is reasonable to conclude that CuO is the major phase present in copper oxide QDs with trace of amounts of sub-oxides (including $Cu_2O$).

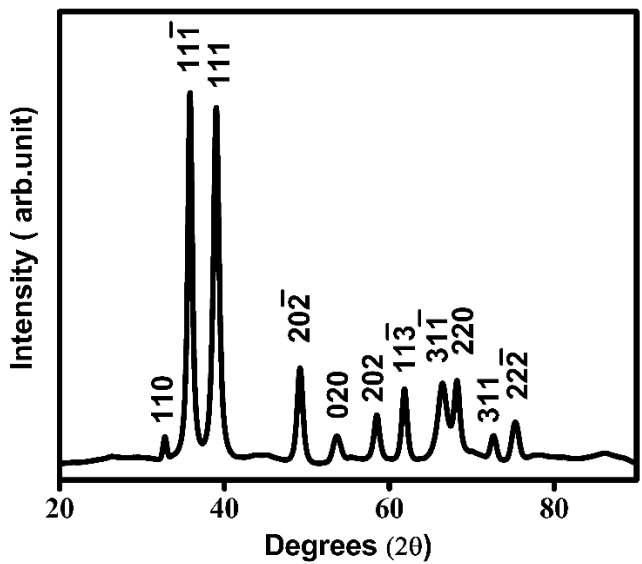

**Figure 1.** XRD pattern of CuO nucleated using higher concentration of precursors (30 mM $Cu(OAc)_2.2H_2O$ and 180 mM NaOH) exhibiting monoclinic phase (Space group: C 2/c,#15).

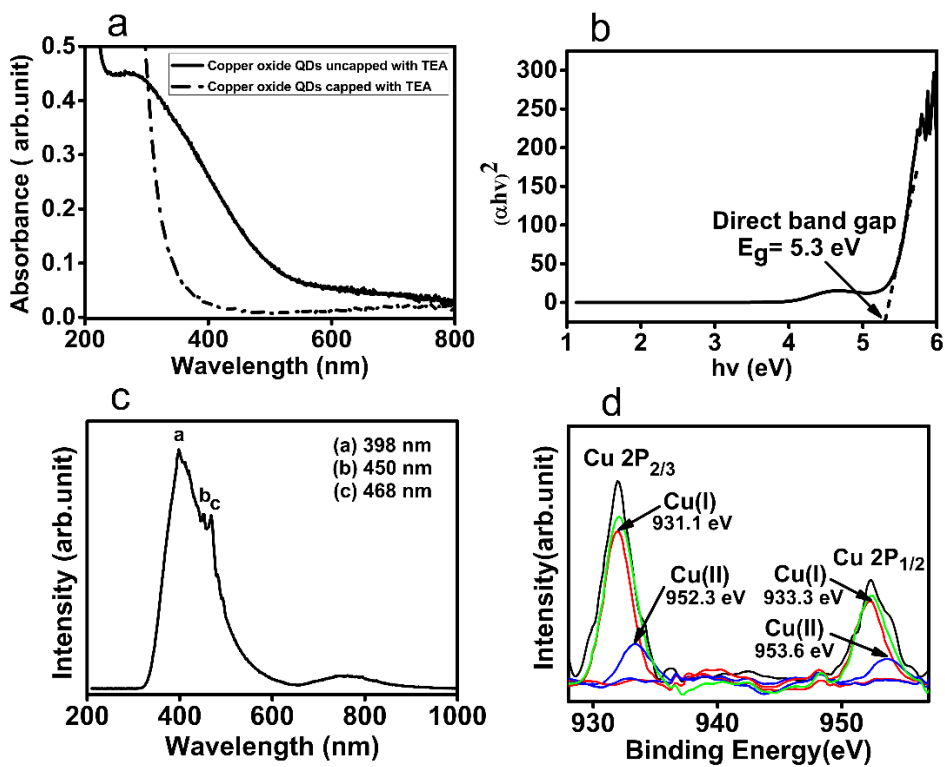

**Figure 2.** (a) UV–Vis absorbance spectra of uncapped and TEA capped copper oxide QDs, (b) Tauc plot for copper oxide QDs, (c) PL spectrum of copper oxide QDs using Xenon lamp of 200 nm as excitation, (d) smoothed and curve fitted XPS spectra of copper oxide QDs; 2p peak deconvolution is done with Lorentzian function. The spectrum obtained indicates that Cu exists in mixed valence state in this material.

Figure 2(a) shows the optical absorption spectra of capped copper oxide QDs and that of uncapped copper oxide QDs. The absorption spectrum clearly shows that the band edge absorption of capped copper oxide QDs is at 370 nm. The capped copper oxide QDs and uncapped copper oxide QDs obtained are blue shifted (4.1 eV) when compared to that of bulk CuO. This is attributed to quantum size confinement. The absorption wavelength of TEA is much lower than copper oxide QDs hence it is reasonable to expect it not to interfere with absorption edge of copper oxide QDs.

Particle size of semiconductor QDs can be calculated from the absorbance spectra by using the Brus' equation. However we observed that Brus' equation fails to predict particle size (as determined by TEM) in case of ultra-small copper oxide QDs. The Brus' equation is written as:

$$E_g = E_{bulk} + \frac{h^2}{8r^2}\left(\frac{1}{m^*_e} + \frac{1}{m^*_h}\right) - \frac{e^2}{4\pi\varepsilon_0\varepsilon_r\gamma_e}$$

In the above equation, '$E_g$' and '$E_{bulk}$' are the band gaps of copper oxide QD and bulk CuO respectively. Also 'h' is the Plank's constant, 'r' is the particle radius and '$\gamma_e$' is the Bohr exciton radius of CuO ( ~ 6.6 nm to 28.72) [32,33], $m^*_e$ and $m^*_h$ are the effective masses of electrons [32,34], and holes [33] in CuO. The values $m^*_e$ and $m^*_h$ are 0.4-0.95 $m_0$ and 7.9 $m_0$ respectively (where $m_0$=mass of an electron). The Bohr exciton radius of CuO is found to lie between 6.6 nm and 28.72 nm. Given the physical size of the capped, ultra-small copper oxide QDs studied here with radius ranging (size ~ 2.4±0.5 nm); it is reasonable to believe that these particles are in the strong confinement regime.

The absorption data from capped copper oxide QDs is used to generate Tauc plots to determine the energy gap of the QDs. The Tauc relation used is the following:

$$\alpha h\nu = A(h\nu - E_g)^n$$

In this equation, '$E_g$' is the band gap of the material, '$\nu$' is the frequency of the incident radiation, 'h' is Planck's constant, '$\alpha$' is the absorption coefficient in $cm^{-1}$, 'A' is a constant related to the material and the matrix element of the transition, and coefficient 'n' depends on the nature of the transition (n = 1/2 for the direct allowed transition or 2 for an indirect transition). Hence for direct band gap materials, the energy gap is determined by plotting '$(\alpha h\nu)^2$' versus 'h$\nu$' and finding the intercept on the 'h$\nu$' axis; this is done by extrapolating the plot to $(\alpha h\nu)^2 = 0$. The best linear relationship is obtained by plotting $(\alpha h\nu)^2$ against photon energy (h$\nu$), indicating that the absorption edge is due to a direct allowed transition in the copper oxide QDs. Figure 2(b) shows the Tauc plot from which the direct band gap is obtained in the range 5.3 eV. Compared with bulk CuO, which has the band gap of 1.2 eV; the ultra-small copper oxide QDs are evidently blue shifted (~5.3eV). Since this blue shift is inexplicable using Brus' equation, we consider this to be an anomalous blue shift. To some extent this result can be reconciled with the report of Borgohain et.al [35], who has reported CuO QDs (~ 10 nm), synthesized using an electrochemical approach, with a band gap of ~4.13 eV . In addition to this Kailash et.al [36], reported CuO QDs (~ 6 nm), synthesized using spray pyrolysis method shown band gap of ~5.5 eV. This too arguably is a case where in the Brus' equation does not hold true. Hence given Borgohain and Kailash's work, and the current report, questions concerning the precise extent of off-stoichometry in copper oxide QDs below a certain size regime, and its impact on its electronic structure are legitimate questions.

PL spectrum of the copper oxide QDs (ref: Figure 2(c)) shows three main broad emission bands centred at 398nm, 450 nm, and 468 nm. The PL peak at 398 nm is related to the band-edge emission of copper oxide QDs; the remaining two peaks at 450 nm, and 468 nm are due to the band edge emission from the new sublevels [37]. These levels are most likely due to the defects present in these copper oxide nanostructures. Figure 2(d) shows the XPS analysis; the peaks obtained at ~931.9 and ~952.3 eV have a significant full width half-maximum. These correspond to the atomic term symbols $2P_{3/2}$ and $2P_{1/2}$ of $Cu^{1+}$ [38]. The peaks obtained at ~933.3 and ~953.6 eV are correspond to $Cu^{2+}$ [38]. This shows that Cu exists in mixed valence state on the surface of this copper oxide QDs.

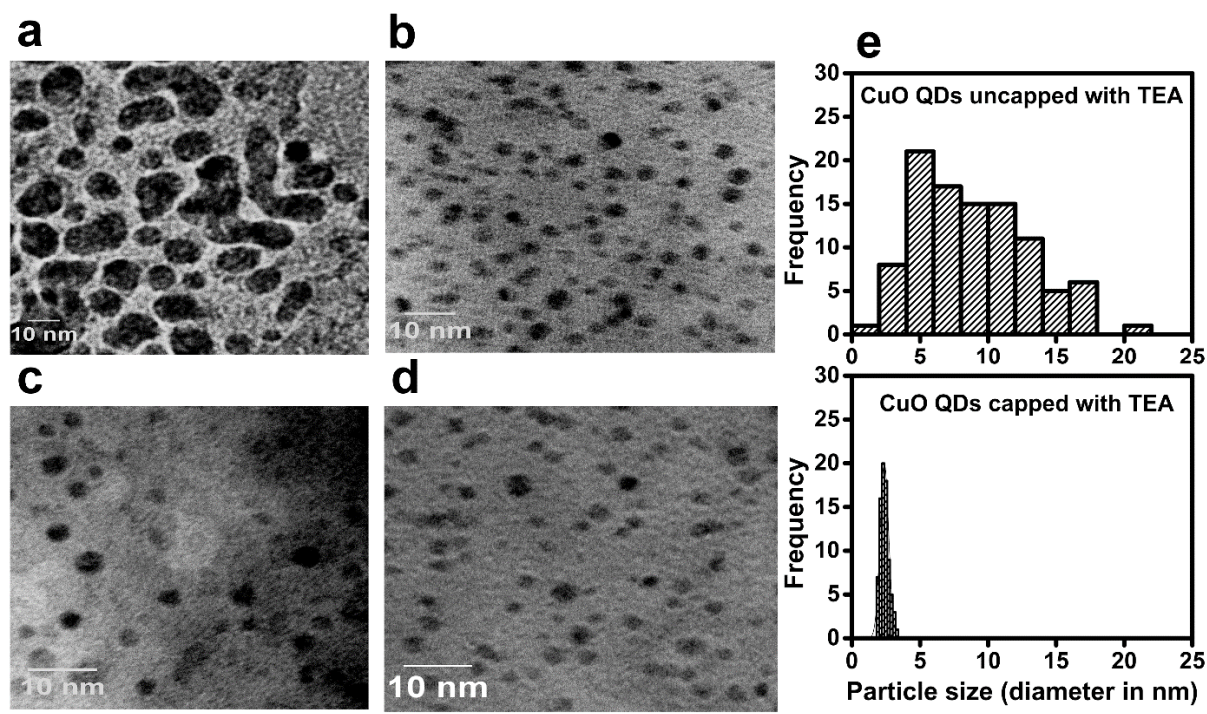

**Figure 3**. TEM images of (a) uncapped copper oxide QDs and (b) capped copper oxide QDs (c) TEM images of digestively ripened copper oxide QDs that are 120 days old. It may be noted from (a) that without capping particle coalescence is a concern. (d) TEM images of digestively ripened copper oxide QDs that are 1 year old. (e) Shows particle size distribution of copper oxide QD before and after digestive ripening.

TEM measurements (ref: figure 3(a) and 3(b)) confirm that both uncapped and capped copper oxide QDs have spherical morphology. The size distribution graph is plotted by using Image-J based analysis; image processing allows evaluation of particle sizes using the TEM images of QDs. We notice that capping with TEA results in reduction in the average particle diameter from 6 ± 4 nm to 2.4 ± 0.5 nm (ref: figure 3(e)); this is due to DR. It may be noted that after DR in ZnO, this is only the second transition metal oxide system in which DR is reported. In order to study the stability of capped CuO QDs, TEM analysis is carried out on the samples that are 100 days older (ref: figure 3(c)), and 1 year older (ref: figure 3(d)), it shows that particles are stable ( >1 year), and show no agglomeration.

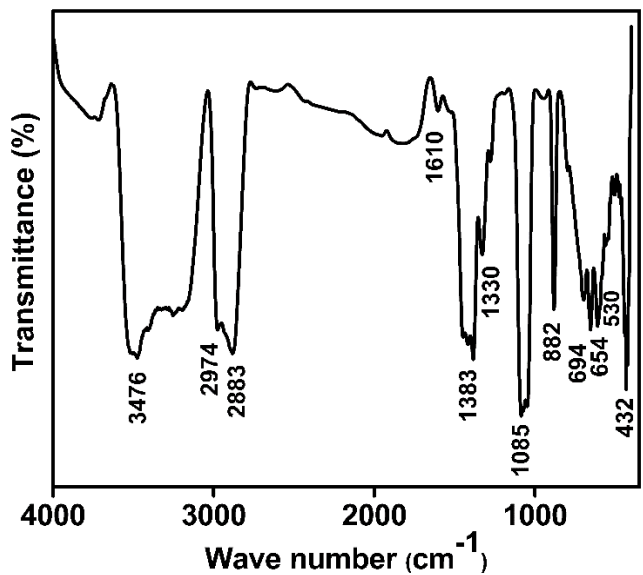

**Figure 4.** FT-IR spectrum of copper oxide QDs

Figure 4. shows the FTIR spectrum of prepared copper oxide QDs; the IR bands centred at 432,530, and 654 $cm^{-1}$ are due to presence of monoclinic copper(II) oxide [34]. These are in fact the only bands that are attributable to CuO. The band centred at 694 $cm^{-1}$ is attributed due to Cu(I)-O bond which result may due to presence of traces of cuprous oxide along with cupric oxide in and on the copper oxide QDs [39]. Borgohain et.al [40], reported CuO QDs by using electrochemical route which exhibits three absorption peaks at around 468, 529 and 590 $cm^{-1}$; however the CuO QDs in the present study are shifted slightly from reported values. This too may be due to the presence of off stoichiometry in the CuO QDs reported. The bands centred at 2883 and 1330 $cm^{-1}$ are due to C-H stretching and bending due to the presence of $CH_3$ group in TEA. On the other hand, the 2974 and 1383 $cm^{-1}$ modes are due to C-H stretching and the bending

of ($CH_2$-). 882 $cm^{-1}$ is due to C-C stretching; the 1085 $cm^{-1}$ peak is due to presence of C-N stretching in TEA. 3476 and 1610 $cm^{-1}$ are stretching and bending vibrations of hydroxyl group, which may be either from moisture or due to traces of copper hydroxide phase.

It has been noticed that the ξ of uncapped copper oxide QDs is 12 ± 2 mV; that of capped copper oxide QDs is 31± 2 mV. The capped copper oxide QDs are more stable (>1 year); this observed stability is due to electrostatic stabilization as well as due to steric repulsions. It is noteworthy that in all these cases, the ξ is positive indicating that the surface is positively charged. This is most likely due to adsorption of positively charged ions on the surface copper oxide QDs. The uncapped copper oxide QDs has positive ξ which is due to adsorption of $H^+$ on its surface, the sources for $H^+$ ions are the water molecules which are present in precursor and water molecules which are formed during metathesis reaction as by-products. In case of capped copper oxide QDs, TEA is physisorbed on its surface and which enhances the adsorption of $H^+$ due to presence of lone pair on the nitrogen atom in TEA. It may be noted that this mechanism (ref: figure S1 in the supplementary information) seems to be operational in case of both ZnO, as well as copper oxide. However these are subtle distinctions due to the (i) multivalency of copper, and (ii) mixed phases present in the reported material.

4. **Conclusions**

Ultra-small, digestively ripened, copper oxide nanoparticles are synthesized by using green solution, soft chemical (chimie douce) approach. In fact this is only the second transition metal oxide system in which DR is reported, so far. These copper oxide QDs are stabilized for (>1 year) in ethanol using triethanolamine as capping agent; a combination of electrostatic and steric factors are responsible for the observed stability. The DR process reported in copper oxide QDs results in very monodisperse (2.4 ± 0.5 nm) particles, with spherical morphology. An anomalous increase in band gap (~5.3 eV), inexplicable using Brus' equation is attributed due to off-stoichiometry and mixed valence seen in the copper oxide QDs reported.

**Acknowledgments**

We thank the Department of Metallurgical and Materials Engineering and Department of Chemistry, Indian Institute of Technology Madras. We would like to thank the Department of Science and Technology, Government of India for support through the project nos. DST FILE NO. YSS/2015/001712 and DST 11-IFA-PH-07. Bhusankar gratefully acknowledges the support from an HTRA fellowship and thanks Professor T. Pradeep, IIT Madras for the XPS experiment. We acknowledge SAIF, IIT Madras, Chennai for using Spectrofluorimeter.

## Supplementary information

**Ultra-small (r<2 nm), stable (>1 year), mixed valence copper oxide quantum dots with anomalous band gap**

**Bhusankar Talluri [1,2], Edamana Prasad[2] and Tiju Thomas *,[1]**

[1] Department of Metallurgical and Material Engineering, Indian Institute of Technology Madras, Chennai-600036, India

[2] Department of chemistry, Indian Institute of Technology Madras, Chennai-600036, India

Correspondence:

*,[1] E-mail: tijuthomas@iitm.ac.in, tt332@cornell.edu; Fax: +91-44- 2257-4752 ; Tel: +91- 44- 2257-5781 (Lab)

Contents:

1. Proposed mechanism for the observed increase in $\xi$ of capped copper oxide QDs

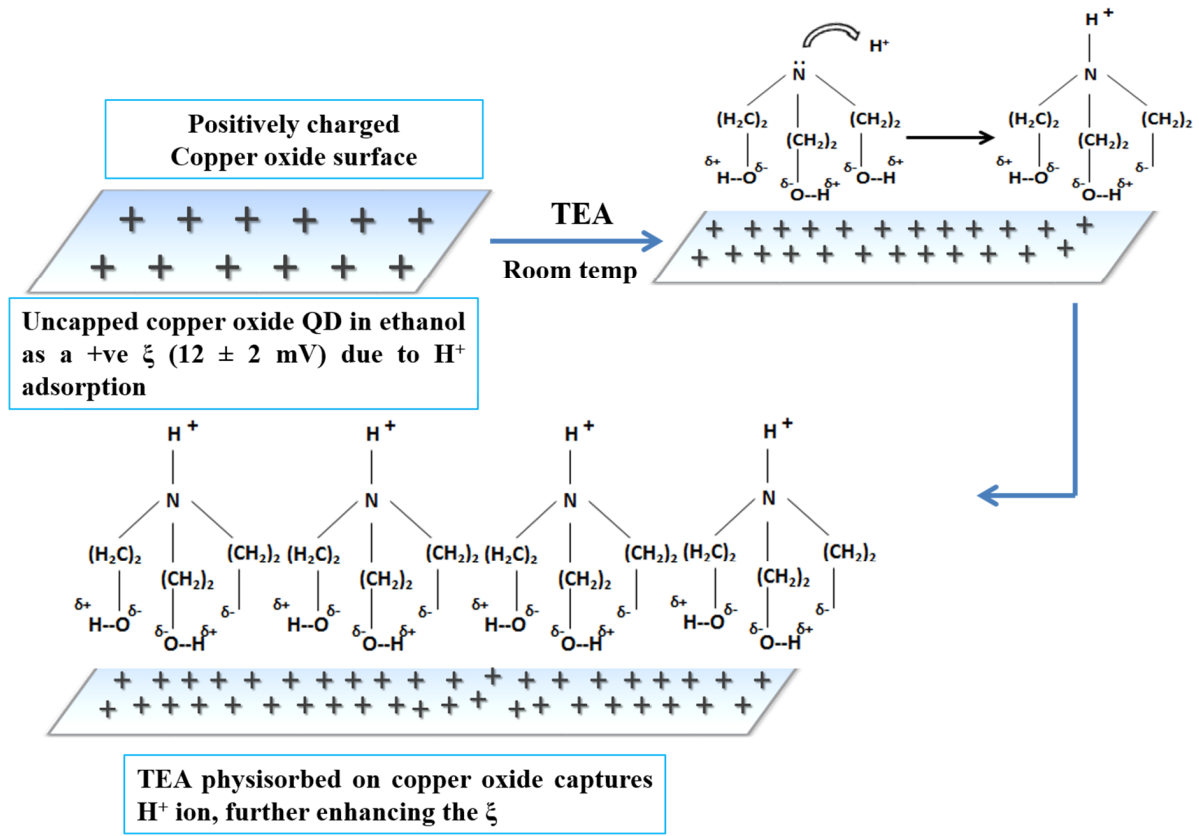

Figure S1. Proposed mechanism for the observed increase in ξ of capped copper oxide QDs (when compared to uncapped copper oxide QDs).

15